\documentstyle[graphicx,12pt,aaspp4]{article}

\begin{document}

\title{WIND INTERACTION MODELS FOR THE AFTERGLOWS OF GRB 991208 
and GRB 000301C}

\author{Zhi-Yun Li and Roger A. Chevalier}
\affil{Department of Astronomy, University of Virginia, P.O. Box 3818}
\affil{Charlottesville, VA 22903}
\affil{zl4h@virginia.edu, rac5x@virginia.edu}


\begin{abstract}

The simplest model of the afterglows of the gamma-ray bursts (GRBs)
envisions a spherical blast wave with a power-law distribution of 
electron energy above some cutoff running into a constant density 
medium. A refinement involves a narrow jet, often invoked to explain 
the steep decline and/or steepening of light curves observed in some 
afterglows. The
constant (ambient) density jet model has been applied  
to GRBs 991208 and 000301C, based to a large extent on radio 
observations. We show that, for these two sources, a spherical 
wind model (with an $r^{-2}$ density ambient medium) can fit the 
radio data as well as the jet model.    
The relatively steep decline and the fairly abrupt steepening of the 
R-band light 
curves of, respectively, GRB 991208 and GRB 000301C can be accounted 
for with a non-standard, broken power-law distribution of electron 
energy. 
Our model predicts a slower late decline for the radio flux than does
the jet model.

%
%
\end{abstract}

\subjectheadings{gamma rays: bursts - stars: mass loss}

\section{INTRODUCTION}

The simplest model of the afterglows of gamma-ray bursts (GRBs) involves 
a relativistic spherical blast wave running into a constant density, 
presumably interstellar, medium (M\'esz\'aros \& Rees 1997; Katz 1994; 
see Piran 1999 for a review). The afterglows are emitted by non-thermal
electrons whose energy distribution is usually assumed to be a 
power-law above some cutoff. The predicted power-law decay of
the afterglow emission with time has been subsequently observed at 
X-ray (Costa et al. 1997), optical
(van Paradijs et al. 1997), and radio (Frail et al. 1997) wavelengths,
giving basic confirmation to this now ``standard'' picture. 

There are observed features of afterglows that are difficult to accommodate 
by the simplest model. One of the most noticeable is the steepening 
of the optical light curves. Good examples include GRBs 990123 (Kulkarni 
et al. 1999), 990510 (Harrison et al. 1999; Stanek et al. 1999), and 
991216 (Halpern et al. 2000). The steepening is usually attributed to
a jet-like, instead of spherical, blast wave (Rhoads 1997;
Sari, Piran \& Halpern 1999). 

Radio data are especially useful for analyzing the afterglows of GRBs
because the self-absorption frequency, $\nu_a$, and the characteristic
frequency of the lowest energy electrons, $\nu_m$, both typically lie
in this range.
GRB 970508 was followed extensively at radio wavelengths (Frail,
Waxman, \& Kulkarni 2000) and Frail et al. (2000) have suggested that
there was an initial spherical relativistic expansion 
in a constant density medium (up to day 25),
followed by lateral jet expansion (days 25--100) and 
spherical non-relativistic expansion (days 100--400).
The same data were modeled by Chevalier \& Li (2000) as
a spherical blast wave running into an $r^{-2}$ density medium 
characteristic of a constant mass loss rate and velocity circumstellar 
wind, possibly of Wolf-Rayet star origin.
The model  missed an early low frequency radio data point and some
infrared points, but it  captured most of the observed behavior over 400 days.
The question of a wind vs. a constant density surrounding medium is crucial
for the question of the progenitors of GRBs, because massive stars,
one of the leading candidates for GRB progenitors (Paczy\'nski 1998),
should be surrounded by a wind.

Fairly extensive radio observations are now available for GRBs 991208 and
000301C (Galama et al. 2000; Berger et al. 2000). At  optical 
wavelengths,
the light curves of both sources are somewhat unusual: those of GRB 991208 
are steeper than $t^{-2}$ (Sagar et al. 2000a), and those of GRB 000301C 
show pronounced steepening (Rhoads \& Fruchter 2000;
Masetti et al. 2000; Sagar et al. 2000b; Jensen et al. 2000). 
These two sources have been modeled as a jet running
into a constant density medium by Galama et al. (2000) and Berger et al. 
(2000), based on both radio and optical data. The authors ruled out the 
simplest, spherical wind model with a standard, power-law electron energy 
distribution. Here, we show that the spherical wind model 
fits the observed radio data as well as  
the constant density jet model. We also demonstrate that a non-standard 
distribution of electron energy can account for both the steep decline 
and the steepening of the optical light curves of, respectively, GRBs 
991208 and 000301C. In \S\ref{model}, we review the spherical wind model 
and describe the non-standard electron energy distribution. The model
is then applied to the GRBs 991208 and 000301C in \S\ref{source}. We 
discuss our results 
in  \S\ref{discuss}. 

\section{WIND MODEL WITH NON-STANDARD ELECTRON ENERGY DISTRIBUTIONS}
\label{model}

Our spherical wind interaction
model was originally developed to explain the radio 
observations of GRB 980425/SN 1998bw (Li \& Chevalier 1999), and has 
been subsequently applied to GRBs 980519 (Chevalier \& Li 1999) and 
970508 (Chevalier \& Li 2000). It involves a (trans-)relativistic blast 
wave propagating into a circumstellar wind. Previously, we have shown 
that the inferred wind 
properties are in the range of those expected from Wolf-Rayet stars. We
assume that the post-shock material is distributed uniformly inside a 
thin shell, and approximately determine the blast wave dynamics  using 
shock jump conditions, and particle and energy conservation. The
dynamics agree with those from the more exact self-similar solutions 
of Blandford \& McKee (1976) to within a factor of order unity. As usual, 
we consider synchrotron radiation as the emission mechanism of afterglows
and assume that a constant fraction $\epsilon_e$ ($\epsilon_B$)
of the total energy goes into the radiating electrons (magnetic field). 
The standard prescription for the energy distribution of electrons is
a power-law above some minimum Lorentz factor, $\gamma_{\rm min}$, 
with a constant power-law index $p$. The model takes into account 
relativistic effects and self-absorption (important at the radio wavelengths), 
but not cooling. The cooling effects can be accounted for approximately
once the cooling frequency $\nu_c$ is estimated (Sari, Piran \& Narayan 
1998; Chevalier \& Li 2000). We adopt a flat universe with Hubble 
constant $H_0=65$ km s$^{-1}$ Mpc$^{-1}$ for cosmological corrections. 

The spherical wind model with the standard, power-law distribution of
electron energy 
works rather well for GRB 970508 (Chevalier \& Li 2000). Applying it to 
GRBs 991208 and 000301C runs into a difficulty: we cannot fit the radio 
and optical data simultaneously; the optical light curves do not decline
with time fast enough. To overcome this difficulty, we are motivated
to seek a non-standard electron energy distribution that steepens at 
high energies. Through trial and error, we find that a wind model 
with the following broken power-law distribution can fit all of the data 
for both sources reasonably well:
\begin{equation}
{d N_e\over d \gamma} = C_1 \gamma^{-p_1}, \hskip 1cm {\rm if}\ \ \ 
	\gamma_{\rm min}< \gamma < \gamma_{\rm b},\break
\label{broken1}
\end{equation}
\begin{equation}
	=C_2\gamma^{-p_2}, \hskip 1cm {\rm if}\ \ \ 
	\gamma > \gamma_{\rm b},\ \ \ \ 
\label{broken2}
\end{equation}
where $\gamma_{\rm b}$ is the break Lorentz factor, $p_1$ and $p_2$ are the
power-law indexes for electrons below and above the break $\gamma_{\rm b}$, 
and the coefficients $C_1$ and $C_2$ are related 
through $C_2=C_1\gamma_{\rm b}^{p_2-p_1}$ to ensure continuity of the 
distribution. We further assume that the ratio
of the break and minimum energies, $R_{\rm b}\equiv \gamma_{\rm b}/
\gamma_{\rm min}$, remains constant in time so that the relative shape
of the distribution is time invariant. Given $p_1$, $p_2$ and 
$R_{\rm b}$, the minimum Lorentz factor $\gamma_{\rm min}$ and the
coefficient $C_1$ are determined from the total number and energy of 
the radiating electrons, which in turn are determined from the dynamic 
evolution of the GRB blast wave. More elaborate prescriptions of the
electron energy distribution may produce better
fits to the observed data, but would introduce more free parameters. 
The implications of the above distribution on particle acceleration 
will be discussed in \S\ref{discuss}.

\section{OBSERVED SOURCES}
\label{source}

\subsection{GRB 991208}
\label{grb1208}


GRB 991208 was first detected with the Interplanetary Network (IPN) by the
spacecrafts {\it Ulysses}, WIND and NEAR on December 8, 1999, at 04:36:52 UT
(Hurley et al. 2000). The main properties of its   optical and
radio afterglows 
are discussed in depth by, respectively, Sagar et al. (2000a) and 
Galama et al. (2000). 
The optical afterglow has one of the most steeply 
declining light curves, with the usual temporal decay index $\alpha\approx 
-2.2$ (Sagar et al. 2000a), assuming that the flux density evolves as 
$F_\nu\propto t^\alpha\nu^\beta$. The spectral index on 
December 16.68, 1999 is determined 
to be $\beta=-0.75\pm 0.03$, based on the observed flux in the K-band 
(Bloom et al. 1999) and extrapolated fluxes in the R- and I-band (Sagar
et al. 2000a), with negligible Galactic extinction. A day earlier 
on December 15.64, the fully calibrated Keck-II spectrum yields an
index $\beta=-0.9\pm 0.15$ between $3850$ \AA\ and $8850$ \AA\
 (Djorgovski et al. 
1999). The radio afterglow was observed extensively at a number of 
frequencies between 1.43 and 250 GHz over a period of two weeks 
(Galama et al. 2000). The multi-frequency data set allows an 
approximate determination by the authors of the evolution of several 
key synchrotron parameters: the absorption frequency $\nu_{a}\propto 
t^{-0.15\pm 0.23}$, the ``typical'' frequency $\nu_{m}\propto t^{-1.7
\pm 0.7}$, and the peak flux density $F_{m}\propto t^{-0.47\pm 0.20}$.
The redshift of the source was determined to be $z=0.706$ (Dodonov et al. 
1999; Djorgovski et al. 1999).

The steep decline of the optical light curves of GRB 991208 resembles those 
of GRB 980326 and especially GRB 980519 (Halpern et al. 1999; Jaunsen
et al. 2000). 
For GRB 980519, we have developed
a spherical wind model with a standard power-law electron energy 
distribution and found that it fits the available data, from radio 
to optical to X-ray, reasonably well (Chevalier \& Li 1999). To account
for the steep light curve decline of GRB 980519 (with $\alpha=-2.05\pm 
0.04$; Halpern et al. 1999), an electron energy power-law index of about 
$p=3.0$ is required. The value of $p$ is higher than 
that found in other GRB afterglows (see Table~1 of Chevalier \& Li 2000) but 
is within the range found in radio supernovae. 
To explain the even steeper decline of the optical light curves
of GRB 991208 (with $\alpha=-2.2\pm 0.1$; Sagar et al. 2000a) using the 
standard spherical wind model, an even higher value of $p\ge 3.3$ would 
be required (Galama et al. 2000). Such a large value of $p$ would be 
in conflict with the value $p=2.52$ determined on December 15.5, 1999
(7.3 days after the burst) from the spectral distribution at several
frequencies ranging from 1.43 GHz to
R-band (Galama et al. 2000). Since the decline in the observed optical
flux (with $\alpha\approx -2.2$) is much steeper than that predicted by 
$p=2.52$ (which corresponds to $\alpha=-1.64$ below the cooling frequency, 
assuming slow cooling), the value of $p$ should be smaller (larger) than 
$2.52$ before (after) day 7.3 to account for the optical flux   
relative to that in the radio. The 
apparent variation required in the value of $p$ motivates us to seek a 
non-standard distribution of electron energy with a non-constant power-law
index $p$. One such non-standard distribution is the broken power-law 
distribution specified by equations (\ref{broken1}) and (\ref{broken2}) 
in \S\ref{model}.

We apply the spherical wind model with a broken power-law electron energy
distribution, as described in \S\ref{model}, to the afterglow of 
GRB 991208. Through trial and error, we find that a ``standard'' model with 
the following combination of parameters 
fits the radio through R-band data reasonably well: the electron energy 
fraction $\epsilon_e=0.03$, magnetic energy fraction $\epsilon_B=0.01$, 
total blast wave energy $E=2.7\times 10^{52}$ ergs, wind mass loss rate
${\dot M}=4.0\times 10^{-6}$ M$_\odot$ yr$^{-1}$
for a wind velocity of $v_w=1,000$ km s$^{-1}$, electron energy 
distribution power-indexes
$p_1=2.0$ and $p_2=3.3$, and the ratio of the break to minimum Lorentz
factors $R_{\rm b}=50$. The model fits are shown in Fig.~1. The radio data are 
taken from Table~1 of Galama et al. (2000) and R-band data from Table~1 
of Sagar et al. (2000a). Note that the above combination of parameters
is not unique. Indeed, there is a family of parameter combinations, with 
$\epsilon_e\propto\epsilon_B^{-1/5}$, $E\propto\epsilon_B
^{-1/5}$ and ${\dot M}/v_w\propto\epsilon_B^{-2/5}$, that have the same 
fitting curves, as long as $\epsilon_B$ is small enough to keep the
cooling frequency above the R-band at the beginning of the afterglow
observations (Li \& Chevalier 1999) and the electron energy distribution 
remains the same. Different electron energy distributions are also 
possible; the parameter $p_2$ that describes the distribution of the 
high energy electrons above the break is particularly ill-constrained
by the light curves shown in Fig.~1 alone. It turns out that a ``variant''  
model with a 
combination of the energy distribution parameters $p_1=2$, $p_2=\infty$ 
(i.e., no high energy electrons above the break at all), and $R_{\rm b}=125$ 
provides an equally good fit to the radio data and a slightly better fit 
to the R-band data. 

Additional constraints come from the spectral index $\beta$ at the optical 
wavelengths. As mentioned earlier, a fully calibrated Keck-II spectrum 
yields $\beta=-0.9\pm 0.15$ on December 15.64. At this time, the 
``standard'' model with $p_1=2$, $p_2=3.3$, and $R_{\rm b}=50$ gives 
$\beta=-1.04$, 
within 1$\sigma$ of the observed value, whereas the ``variant'' model 
with $p_1=2$, $p_2=\infty$ and $R_{\rm b}=125$ gives $\beta=-1.27$, about 
2$\sigma$ away from the observed value. It therefore appears that $p_2$
should be close to $3.3$. A potential difficulty is the value of $\beta
=-0.75\pm 0.03$ on December 16.68, inferred from the observed flux in the 
K-band (Bloom et al. 1999) and {\it extrapolated} fluxes in the R- and 
I-band (Sagar et al. 2000a). However, with only two usable data points in 
the I-band (Sagar et al. 2000a) and possible deviation from a pure power-law 
decay in the {\it single} K-band measurement (Bloom et al. 1999; see R-band 
data for reference), we regard the index inferred on December 16.68 as less
reliable than that measured simultaneously over a wide range of frequencies 
a day earlier. Therefore, we believe that the overall fit to both the radio 
and optical data is acceptable, and suspect that the fit may be improved 
with more elaborate prescriptions of the electron energy distribution than 
the simple broken power-law adopted here. 
%
%
%

Panel (a) of Fig.~1 should be compared with Fig.~1 of Galama et al. (2000),
where they fit the same radio data with a spectral form given by Granot 
et al. (1999a,b), based on the Blandford-McKee self-similar solution of 
a spherical, ultra-relativistic blast wave propagating in a constant 
density medium, and the best-fit power law time evolution of
$\nu_a$, $\nu_m$, and the peak flux $F_{m}$.
 Inspection by eye reveals that the two model fits are of 
comparable quality; both models fit the three highest frequency light
curves rather well. The fits to the three lowest frequency light curves are
less satisfactory in {\it both} models, presumably because the lower frequency 
data are affected more by interstellar scintillation (ISS; Goodman 1997; 
Galama et al., in preparation). It is partially based on 
the low frequency data, which fix the self-absorption frequency $\nu_{a}$, that 
Galama et al. (2000) ruled out the spherical wind model for this source. 
They inferred that  $\nu_{a}$ evolves as $t^{-0.15
\pm 0.23}$, which is incompatible with the evolution, $\nu_{a}\propto 
t^{-3/5}$, predicted by a wind model. The fitted slope, $-0.15\pm 0.23$, 
hinges, however, to a large extent on the { non-detection} on December 
22.96 at 1.43 GHz (see the lower-right panel of their Fig.~2), the lowest 
observing frequency that is most affected by ISS. The poor quality of the
power-law fit to the inferred values of $\nu_{a}$ is reflected in the 
relatively large {reduced} chi-square: 
$\chi^2_{\rm r}=3.5$ for 2 degrees of freedom (see Table 3 of 
Galama et al. 2000). 
Indeed, if one were to 
ignore the value of $\nu_{a}$ based on the non-detection 
(the last data point 
in the leftmost panel of their Fig.~4), the remaining three points would 
yield a much steeper slope. Therefore, this piece of evidence against 
the wind model is not compelling in our opinion. As noted by Galama et al.
(2000), the inferred evolution for two other key parameters, $\nu_{m}
\propto t^{-1.7\pm 0.7}$ and $F_{m}\propto t^{-0.47\pm 0.20}$, is
compatible with the wind model. A second objection raised by 
Galama et al. (2000) was that the spherical wind model requires an unusually 
large electron energy index, $p\ge 3.3$, to account for the steep decline 
of the optical light curves. This would indeed be the case if the electron 
energy distribution were a single power-law, as commonly assumed. The objection
motivated us to seek a non-standard, broken power-law distribution of 
electron energy which, as demonstrated by Fig.~1, explains both the radio 
and R-band data reasonably well. 

Based on the same set of optical and radio data, Galama et al. (2000) reached 
a different conclusion. They favored a constant 
density jet model for GRB 991208. As noted by these authors, the afterglow 
properties inferred from observations differ from the {\it asymptotic} 
model predictions (Sari, Piran \& Halpern 1999) on two accounts. First, the 
inferred decrease of the peak flux density with time, $F_{m}\propto 
t^{-0.47\pm 0.20}$, is substantially slower than the predicted $F_{m}
\propto t^{-1}$. Second, the inferred electron energy index $p=2.52$ on
December 15.5 is larger than the predicted $p\approx 2.2$, based on the 
temporal decay index of $\alpha\approx -2.2$ in R-band. The authors argued 
that a slow transition to the fully 
asymptotic regime (Kumar \& Panaitescu 2000; Moderski, Sikora, \& Bulik
2000) may resolve these discrepancies, 
since during the transition the rate of the peak flux decay {\it should} be 
between that of a spherical model, $F_{m}\propto t^{0}$, and that 
predicted asymptotically, $F_{m}\propto t^{-1}$, and the decay index 
$\alpha$ of the R-band light curve {\it should} be greater than $-p$ 
(or --2.52). While this explanation is  plausible,  
it is far from being proven; detailed modeling of the transition 
is required. 

The continued evolution of the radio flux provides a test of the models.
Galama et al. (2000) note that in their model $\nu_m$ should pass
8.46 GHz at $\sim$12 days and the flux should then decay rapidly,
$F_{\nu}\propto t^{-2.2-2.5}$.
In our model, the flux evolution at this frequency
should tend to $F_{\nu}\propto t^{-1.25}$
({see Fig.~1}). Continued monitoring of the source was apparently 
undertaken (Galama et al. 2000).

\subsection{GRB 000301C}

GRB 000301C was first detected with the RXTE All Sky Monitor and the IPN 
spacecrafts {\it Ulysses} and NEAR on March 01, 2000, at 09:51:37 UT 
(Smith, Hurley \& Cline 2000). The IR/optical properties of its 
afterglow are discussed in Rhoads \& Fruchter (2000), Masetti et al. 
(2000), Sagar et al. (2000b), and Jensen et al. (2000). Two features 
of the IR/optical light curves stand out: (a) large amplitude, short 
time scale variations and (b) a relatively abrupt steepening from a 
temporal decay index of $\alpha\approx -1$ to roughly $-2.7$. The 
spectral index $\beta$ was determined in the IR/optical spectral 
region at several times between day 2 and 14, with values ranging from 
$\sim -0.5$ to $\sim -1.5$ (Rhoads \& Fruchter 2000; Sagar et al. 2000b; 
Jensen et al. 2000). The redshift of the source was determined to be 
$2.03$ (Smette et al. 2000; Castro et al. 2000; Jensen et al. 2000). 
Radio light curves of varying degree of coverage are available at 
four frequencies (4.86, 8.46, 22.5 and 250 GHz; Berger et al. 2000), 
which allow for detailed modeling. 

As pointed out by Berger et al. (2000), the steepening of the IR/optical
light curves rules out the simplest spherical wind model {\it with a 
standard, power-law distribution of electron energy} for this source. A 
spherical wind model with a non-standard distribution of electron 
energy is still possible. Indeed, the model with a simple broken 
power-law distribution, outlined in \S\ref{model}, can fit the IR/optical 
light curve steepening, as well as the radio data reasonably well. This 
is demonstrated in Fig.~2, where we compare the observed radio and R-band 
light curves with those predicted from a wind model with $\epsilon_e=0.04$, 
$\epsilon_B=0.01$, $E=2.3\times 10^{52}$ ergs, ${\dot M}/v_w=4.5\times 
10^{-6}$ M$_\odot$ yr$^{-1}$/$10^3$ km s$^{-1}$, $p_1=2.2$, $p_2=15$, and 
$R_{\rm b}=140$. 
Inspection of panel (b) reveals that the model fits rather well the overall 
steepening of the R-band light curve, but not the short time scale
variations, most noticeably the bump around day 4 and, to a lesser 
extent, the bump around day 7. The first bump has been interpreted
by Garnavich, Loeb \& Stanek (2000) as due to microlensing, although 
it could also be caused by a local enhancement in the ambient density
(Berger et al. 2000) or an impulsive energy input (Panaitescu, 
M\'esz\'aros \& Rees 1998; Li \& Chevalier 1999; Sari \& 
M\'esz\'aros 2000; Dai \& Lu 2000). 
As noted by Berger et al. (2000), the physical process responsible 
for the first bump on the R-band light curve may also explain
the factor-of-two discrepancy between the 250 GHz data taken 
around the same time and the model predictions (see their Fig.~2 
and panel [a] of our Fig.~2). 
Panel (a) of Fig.~2 should be compared with Fig.~2 of Berger et al. 
(2000), where the same radio data set is fitted with a constant density 
jet model. Judging by eye, we find the quality of the two model fits 
comparable: the jet model fits the 250 and 22.5 GHz data slightly better, 
whereas the wind model fits the 8.46 and 4.86 GHz data slightly better. 
We note again that the combination of 
parameters listed above is not unique. In particular, the power-law
index $p_2$ for the high energy electrons above the break $\gamma
_{\rm b}$ is not well constrained. A relatively large value of $p_2$ 
is required nevertheless, to explain the rapid decline of the R-band
flux at late times (with a decay index of $\alpha\approx -2.7$) in the 
context of the spherical wind model. 

Berger et al. (2000) ruled out the possibility that the steepening in 
the light curves of GRB 000301C is due to a time-varying $p$, based 
on the fact that the single value, $p=2.70$, that they inferred 
from a global fit to the radio/IR/optical data set, appears to fit the 
spectral flux distributions simultaneously at both day 4.26 and 12.17 
reasonably well. Since their fitting is based on a model that assumes 
a {\it single} power-law distribution for the electron energy (i.e., a 
{\it constant} value of $p$ at any given time), it does not necessarily 
rule out a curvature in the {\it slope} of the energy distribution 
(i.e., a $p$ {\it that varies with energy} at any given time) as the 
cause for the light curve steepening. Indeed, there is some evidence 
for the curvature in the spectral energy distributions (SEDs) in the
near IR to near UV region compiled by Rhoads \& Fruchter (2000; the left 
panel of their Fig.~3). These SEDs are reproduced in panel (a) 
of our Fig.~3, with the modest amount of Galactic extinction  
($E_{B-V}=0.053$; Schlegel, Finkbeiner \& Davis 1998) already 
corrected for. Note, however, that the data at different frequencies are 
not taken simultaneously and considerable 
uncertainties are involved in interpolating their fluxes to a common 
time for this source with known large amplitude, short time scale 
variabilities. The relatively shallow distribution at day 7.59 could, 
for example, be due to the previously mentioned, second (smaller) bump 
on the R-band light curve around day 7, which may or may not be present 
in the K$^\prime$-band (where the data points are too sparse to tell). 
Nevertheless, there appears to be a general trend that the SED steepens
with time (see also column 5 of Table 3 of Rhoads \& Fruchter). We 
expect such a steepening to occur in our model, as
electrons with higher and higher energy emit into the frequency band. 
The spectral steepening is a robust signature of the curvature in 
the electron energy distribution and could in principle be used 
to distinguish our spherical wind model from the jet model. In panel 
(b) of Fig.~3, we plot the predicted SEDs of the wind model of 
GRB 000301C shown in Fig.~2. As expected, the SEDs steepen with 
time, although they are generally shallower than their observed 
counterparts in panel (a). A better agreement is reached if we invoke 
a modest amount of host galaxy extinction of $A_{\rm V}=0.09$ of the Small 
Magellanic Cloud (SMC) type (Pei 1992), as recommended by Rhoads \& 
Fruchter (2000). A somewhat different value of $A_{\rm V}=0.14$,
also of the SMC-type, is favored by Jensen et al. (2000) on day 3. 

An alternative model for the afterglow of GRB 000301C was proposed by 
Kumar \& Panaitescu (2000b). These authors attribute the steepening of the 
R-band light curve to a sudden, large drop in the density of the 
ambient medium into which the GRB blast wave propagates. Initially, the 
light curve steepens continuously as the blast wave expands freely. It 
tends to a constant decay rate when the observed flux is dominated by 
the emission from the high latitude parts of the blast wave away from
the line of sight to the explosion center. This model and ours are 
similar in that both invoke a sudden change in either the ambient density 
or the electron energy distribution and that the finite light-travel time 
effect plays an important role in shaping the observed light curves. A 
potential problem with the model of Kumar \& Panaitescu is that it predicts
a rather steep flux decay of $t^{-2.8}$ for frequencies above $\nu_{\rm m}$ 
but below $\nu_{\rm c}$; the late-time decay of the observed radio flux at 
8.46 GHz does not appear to be that steep (see Fig.~2), although additional
observations at even later times are needed to draw a firmer conclusion.

Finally, we note that in our model even though the change in the 
power-law index $p$
of the electron energy distribution is discontinuous (jumping from $p_1$ 
to $p_2$), the variation of the spectral index $\beta$ with time is 
smooth (see Fig.~3b). This makes inferring the instantaneous value
of $p$ from the value of $\beta$ at any given time, as has been done  
by, e.g., Sagar et al. (2000b), difficult. 

\section{DISCUSSION}
\label{discuss}

Our study of these two GRBs  can be separated into the radio evolution
and the optical/IR evolution.
ISM + jet models have been proposed for the radio emission from
both sources, but we find that
spherical wind interaction models can produce fits of comparable
quality.
The same is true for GRB 970508 (Frail et al. 2000; Chevalier \& Li 2000).
In the case of GRB 991208, the approximate power law behavior observed
over a factor 10 in time is intermediate between that expected
for a spherical explosion in
a constant density medium and the asymptotic lateral expansion of
a jet (Galama et al. 2000).
Kumar \& Panaitescu (2000a) found that
the transition to the asymptotic jet evolution should take at least
an order of magnitude, perhaps several orders of magnitude, in observer
time.
Yet the model of Berger et al. (2000) for GRB 000301C assumes a rapid
transition to the asymptotic jet evolution.
The uncertainties in the jet models give them additional parameters that
are not present in the wind interaction models
which assume spherical symmetry throughout the evolution.
Kumar \& Panaitescu (2000a) found that for jet expansion in a wind medium,
any break in the observed light curves would be unlikely to be detected,
even for a narrow jet with an opening angle of about a few degrees.
The spherical assumption, which we make,
 should be adequate even if a jet is present.
 
The problem with the simplest wind model is that the optical light
curves of both GRB 991208 and GRB 000301C drop more rapidly than
predicted.
We have solved the problem by invoking a  steepening of the
electron spectrum.
We attribute the steepening to the acceleration process, although
the details of how this occurs are not clear.
Bednarz \& Ostrowski (1998) have studied acceleration in relativistic
shock waves and found that at high shock Lorentz factors, the spectrum tends
to  a power law with $p\approx 2.2$.
However, at low Lorentz factors ($\gamma\approx 3$), which are characteristic
of afterglow shock waves, there are a number of possibilities, including
$p < 2$, depending on the shock parameters.
A particle spectrum that is this flat must steepen at high energy in
order to have finite total energy.
The spectral steepening with $p$ increasing to a relatively large 
value of $3.3$ or more that we propose for GRBs 991208 and 000301C 
would be clearer if X-ray observations of the sources were available. 
Unfortunately, these two sources are not constrained by X-ray 
observations. Spectral steepening with $p$ increasing to a smaller
value may also be possible. In such a case, the X-ray flux would be 
less affected. 

The difference between an intrinsic spectral break and a laterally expanding
jet is that, in the latter case, the break in a light curve should
be achromatic.
It appears to be difficult to decide this point solely on the basis of
optical/IR observations because of the faintness of the sources and
the sparse sampling.
The difference between the radio and optical evolution might be clearer.
The late radio light curve of GRB 000301C is most complete at 8.46 GHz
and the wind model appears to give a better representation of these
data than the jet model (compare Fig. [2] to Fig. [2] of Berger et al. 2000),
although the uncertainties are too large to draw any firm conclusions.
A similar test may be possible for GRB 991208 (see \S~3.1).
Another test provided by radio observations is at low frequencies.
In the wind interaction model, self-absorption is high at early times
so that the low frequency flux should be low.
Both GRB 970508 and GRB 000301C have one early 1.4 GHz radio observation that
is higher than expected in the wind interaction model (Chevalier \&
Li 2000; Fig. [2]); the 1.4 GHz observations of GRB 991208 show a decrease
that is not expected in any model.
More extensive early, low frequency observations would provide
a useful test of  models.

\acknowledgments

We thank J. Rhoads for providing us with the SED data of GRB 000301C 
shown in Fig.~3a. This work was supported in part by NASA grant 
NAG5-8232.

\newpage
\begin{figure}
\includegraphics[scale=0.8]{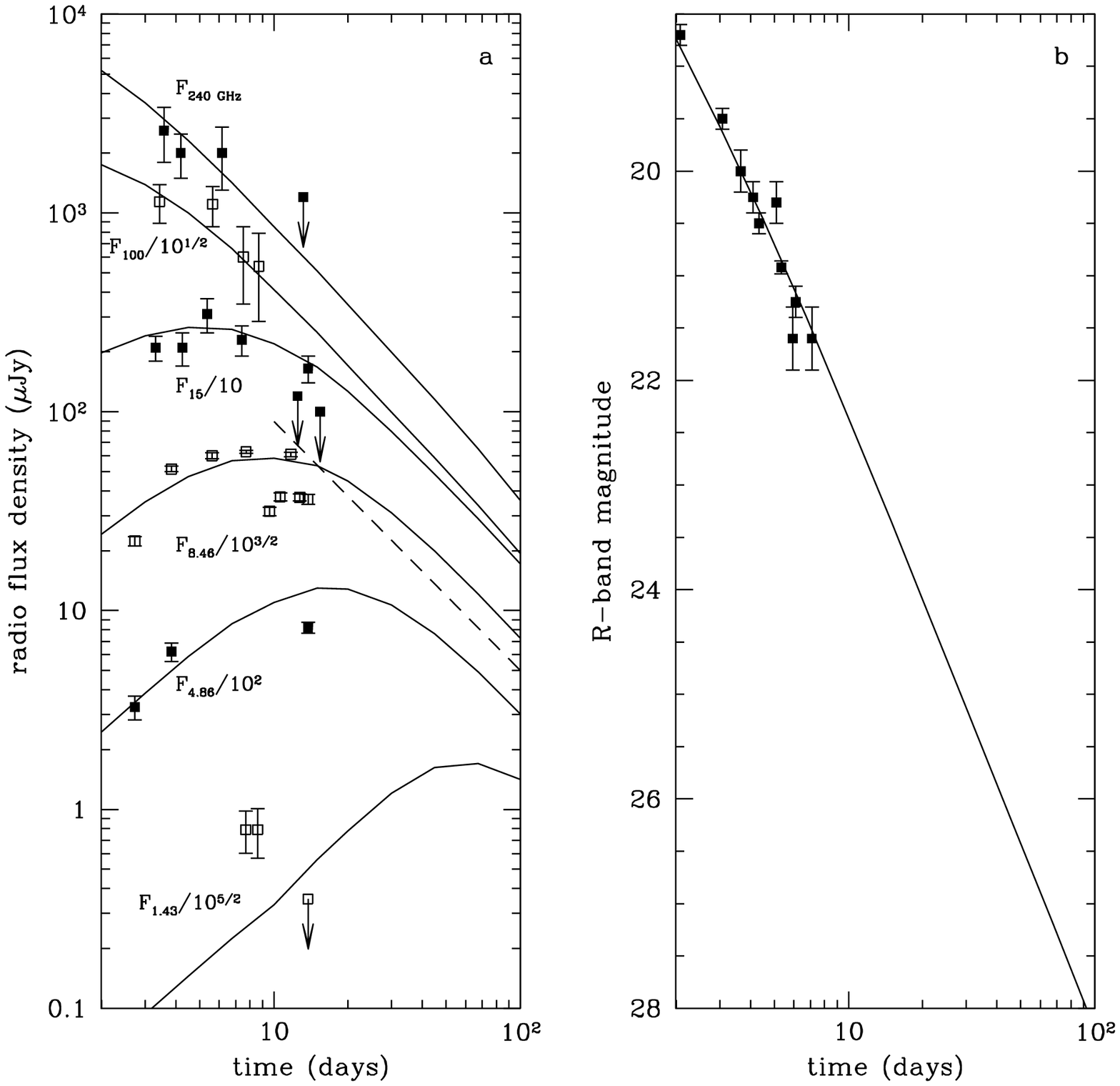}
\caption{The fits of a spherical wind model with a broken power-law electron
energy distribution to the observed (a) radio and (b) R-band data of 
GRB 991208. The model is described in \S\ref{model} and the model parameters 
in \S\ref{grb1208}. The radio data at different frequencies are displaced 
relative to one another to avoid overlap. Curves beyond about day 10
are the model predictions that can be tested when more radio data become 
available (Galama et al., in preparation). {Note in particular
that the predicted 
flux decay at 8.46 GHz tends to a relatively shallow power-law of 
$t^{-1.25}$ (corresponding to $p=2$ in a wind model; dashed line) 
before steepening to $t^{-2.2}$ (corresponding to $p=3.3$). The 
steepening occurs after day 10$^2$, and is not shown.}
}
\end{figure}

\newpage
\begin{figure}
\includegraphics[scale=0.8]{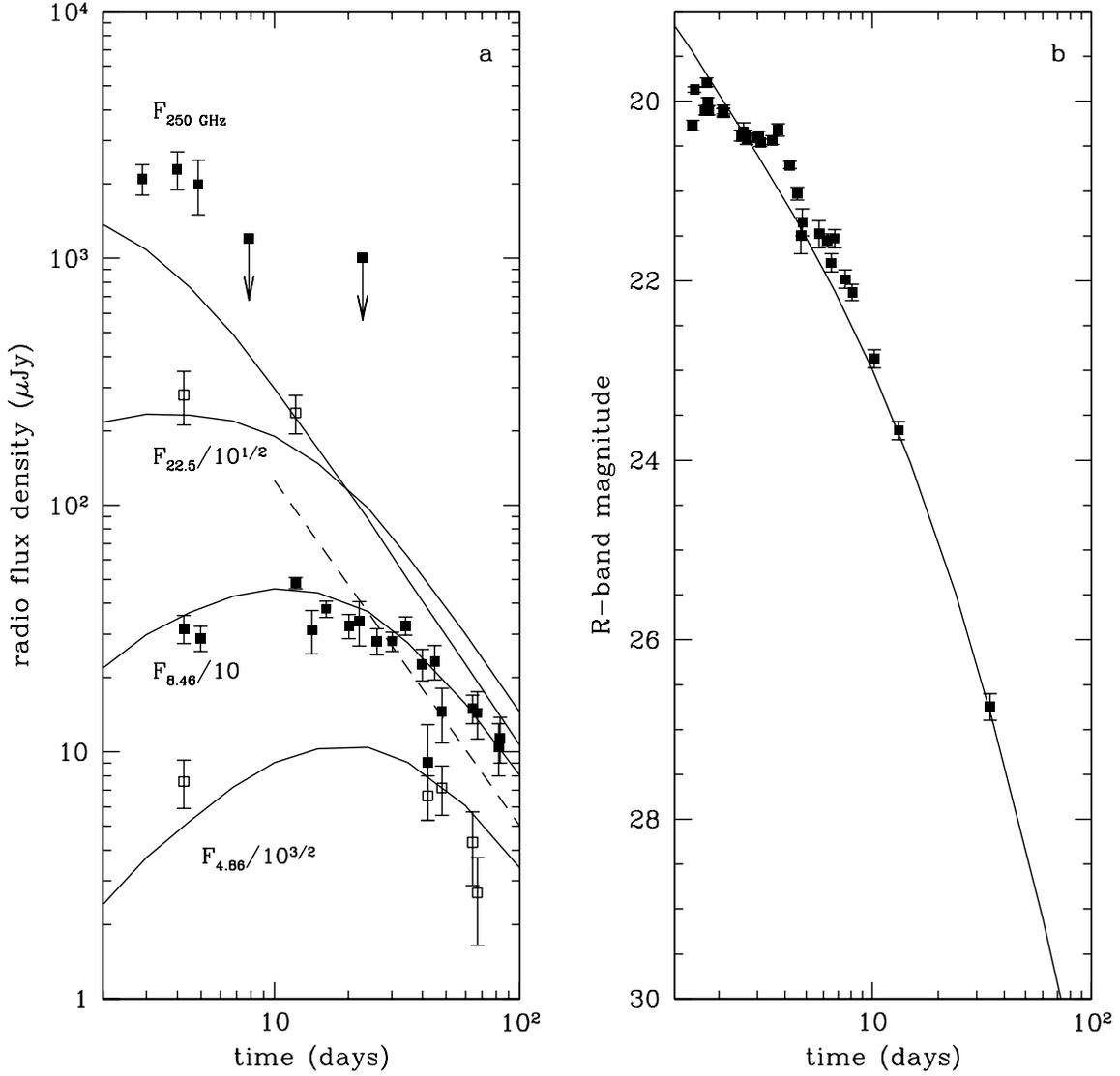}
\caption{Model fits to the observed (a) radio and (b) R-band light 
curves of GRB 000301C. The radio data at different frequencies 
are displaced relative to one another to avoid overlap. A 
power-law of $F_\nu \propto t^{-1.4}$ (corresponding to $p=2.2$ 
in a wind model; dashed line) is plotted for reference. The R-band 
data are increased over the observed values by $15\%$ to account 
for the Galactic extinction ($E_{B-V}=0.053$; Schlegel, Finkbeiner 
\& Davis 1998). } 
\end{figure}

\newpage
\begin{figure}
\includegraphics[scale=0.8]{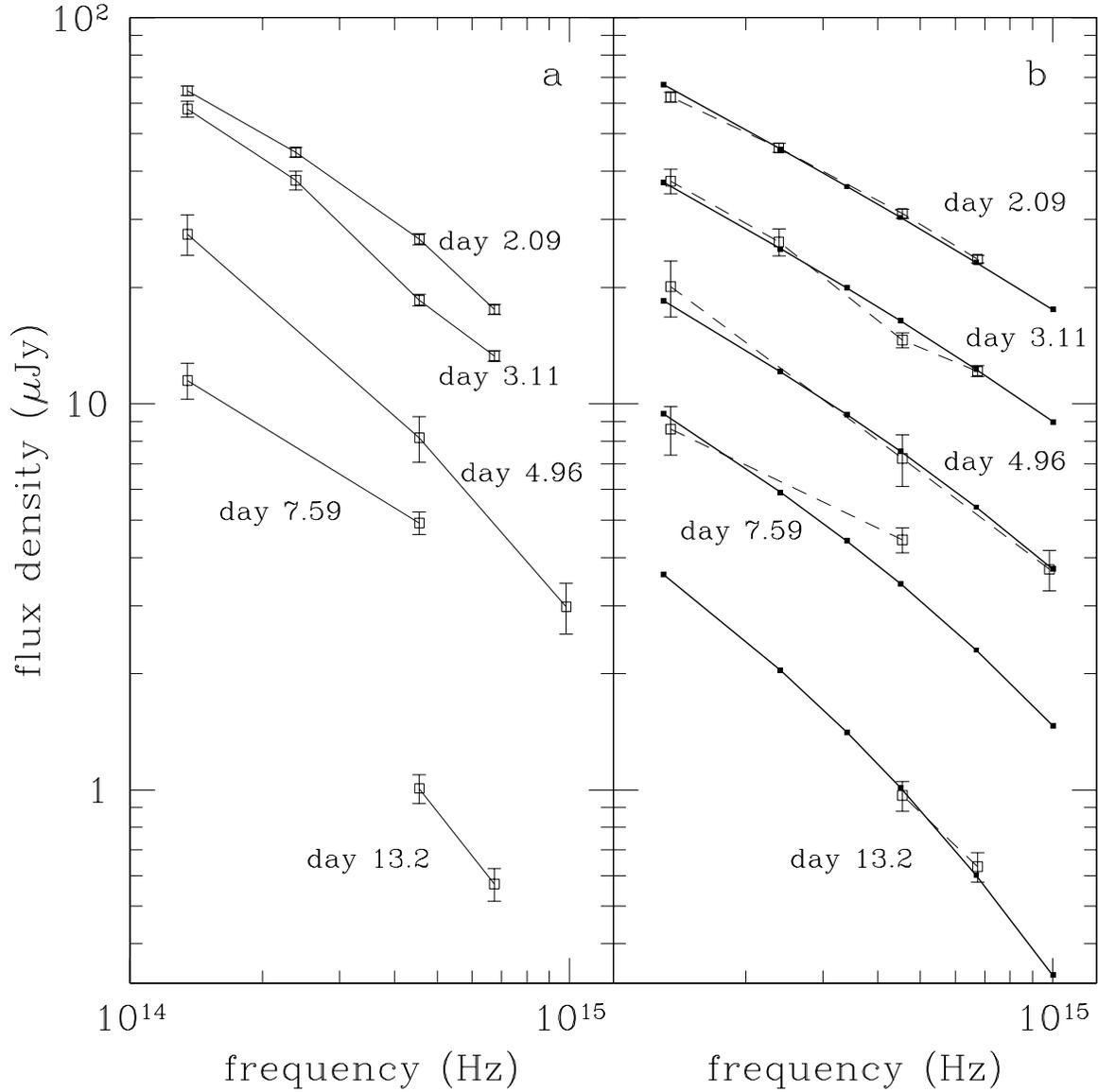}
\caption{The observed (a) and predicted (b) spectral energy distributions 
of the afterglow of GRB 000301C at several epochs. The observed 
distributions in (a) are reproduced from Rhoads \& Fruchter (2000), 
with the Galactic extinction already corrected for. The predicted 
distributions in (b) (solid lines) are generally shallower than their 
counterparts in (a). A better agreement is achieved when an additional 
host galaxy extinction of $A_{\rm V}=0.09$ of the SMC-type is corrected 
for (dashed lines; Rhoads \& Fruchter 2000). We have adjusted the height 
of the dashed lines to allow for a better comparison of the slopes of
the observed and predicted SEDs. 
} 
\end{figure}

\end{document}